\documentclass[twocolumn,superscriptaddress,preprintnumbers,amsmath,amssymb]{revtex4}

\usepackage{SIunits}
\usepackage{pifont}
\usepackage{ulem}
\usepackage{color}
\usepackage{graphicx}

\begin{document}

\title{Strongly Modified Plasmon-Matter Interaction with Mesoscopic Quantum Emitters}

\author{Mads Lykke Andersen}
\email{mlya@fotonik.dtu.dk}
\affiliation{%
DTU Fotonik, Department of Photonics Engineering, Technical University of Denmark, Building 345V, 2800 Kongens Lyngby, Denmark
}%
\author{S{\o}ren Stobbe}
\affiliation{%
DTU Fotonik, Department of Photonics Engineering, Technical University of Denmark, Building 345V, 2800 Kongens Lyngby, Denmark
}%
\author{Anders S{\o}ndberg S{\o}rensen}
\affiliation{%
QUANTOP, Danish Quantum Optics Center and Niels Bohr Institute,
DK-2100 Copenhagen \O, Denmark
}%
\author{Peter Lodahl}
\email{pelo@fotonik.dtu.dk}
\affiliation{%
DTU Fotonik, Department of Photonics Engineering, Technical University of Denmark, Building 345V, 2800 Kongens Lyngby, Denmark
}

\date{\today}


\begin{abstract}
Semiconductor quantum dots (QDs) provide an essential link between light and matter in emerging fields such as light-harvesting \cite{nmat_Polman,Nozik2002115}, all-solid-state quantum communication \cite{PRL_Fattal_Yamamoto_2004}, and quantum computing \cite{Nature_Ladd}. QDs are excellent single-photon sources \cite{P.Michler12222000} and can store quantum bits for extended periods \cite{PhysRevLett.90.206803} making them promising interconnects between light and matter in integrated quantum information networks \cite{PRA_Loss_DiVincenzo_1998}. To this end the light-matter interaction strength must be strongly enhanced using nanophotonic structures such as photonic crystal  cavities \cite{Hennessy}  and waveguides \cite{PhysRevLett.101.113903} or plasmonic nanowires \cite{chang:053002,Akimov2007,singlephotontransistor,PhysRevB.78.153111}. So far it has been assumed that QDs can be treated just like atomic photon emitters where the spatial properties of the wavefunction  can be safely ignored. Here we demonstrate that the point-emitter description for QDs near plasmonic nanostructures breaks down.  We observe that the QDs can excite plasmons eight times more efficiently depending on their orientation due to their mesoscopic character. Either enhancement or suppresion of the rate of plasmon excitation is observed depending on the geometry of the plasmonic nanostructure in full agreement with a new theory. This discovery has no equivalence in atomic systems and paves the way for novel nanophotonic devices that exploit the extended size of QDs as a resource for increasing the light-matter interaction strength.
\end{abstract}

\maketitle

An essential advantage of all-solid-state emitters compared to, e.g., atomic emitters or  molecules, is that they can be positioned deterministically and remain stationary \cite{Springer_Schmidt}. This makes QD-based nanophotonic devices a promising technology for scalable many-qubit systems \cite{O'brien09}. The term "artificial atoms" has been coined to QDs due to their discrete energy levels and their assumed atom-like interaction with light. Presently it becomes clear that QDs in nanostructures lead to a number of surprises distinguishing them from atomic systems, including the recent observations of very broadband radiative coupling in cavity QED \cite{Hennessy} and self-tuning of QD nanolasers \cite{strauf:127404}.  In this Letter we present the first experimental observation and the theoretical explanation of a novel mechanism to enhance the interaction between light and matter induced by the mesoscopic size of QDs. It gives rise to a strongly modified radiative decay that is tailored by the size and shape of the QD electron-hole wavefunctions and can be used as a resource to dramatically enhance the coupling of QDs to plasmonic nanostructures.  The efficient coupling of single emitters to plasmonic nanostructures is currently being investigated intensely for various applications within nanophotonics and quantum optics \cite{nphoton-plasmonicsreview,nmat-brongesma}  enabling highly efficient single-photon sources \cite{chang:053002,Akimov2007,naturephys_spp_duality}, single-photon transistors \cite{singlephotontransistor}, and subwavelength plasmon lasers \cite{spaser_first,Noginov2009}. In all these applications it is essential to understand and enhance the interaction between light and matter, which is the essence of the work presented here.

Figure 1a illustrates the physical system under consideration: QDs are placed at a distance $z$  below a metallic mirror and the electromagnetic field associated with the surface-plasmon polariton resonance at the metal surface  is varying over the extension of the QDs. The QDs are standard-sized  $(\sim 20\times 20 \times 6 \ \nano\meter\cubed)$  clusters of indium arsenide (InAs) embedded in gallium arsenide (GaAs) fabricated by  molecular beam epitaxy. After excitation, the QDs trap single electrons and holes (Fig.~1b), which recombine through different channels with the following rates: excitation of plasmons $\gamma_\mathrm{pl}$, spontaneous emission of photons $\gamma_\mathrm{ph}$,  non-radiative losses in the metal $\gamma_\mathrm{ls}$,  or intrinsic non-radiative recombination in the QDs $\gamma_\mathrm{nr}$,  see Fig.~1c. The impact of the mesoscopic QD size on the radiative coupling to plasmonic nanostructures can be precisely assessed by employing a nanophotonic structure with well-understood optical properties. Here we employ a silver mirror with QDs positioned at precise distances from the surface, whereby the effects of emitter and environment can be unambigously separated. This is not possible in complex structures like photonic crystals \cite{nat04_lodahl_vanmaekelbergh_vos} or plasmonic nanowires \cite{Akimov2007}.

We have measured the decay rate of QDs  ($\gamma_\mathrm{QD}$) versus distance to the silver mirror \cite{JLumin_Drexhage_1970}, see Fig.~2, allowing to distinguish the various decay rates discussed above, see Supplementary Information for further details. Investigating two different orientations of the QDs relative to the silver mirror allows us to unambiguously prove the breakdown of the point-emitter description, i.e. the so-called dipole approximation, which is found to be excellent for atoms, ions, and molecules. In the first sample (the direct structure), the apex of the QDs points towards the silver mirror, while in the second sample (the inverted structure) it points away, see insets of Fig.~2. A point-dipole source would radiate identically in the direct and inverted structures, and the expected decay rate for such an emitter is shown in Fig.~2. We observe that for short distances to the silver mirror the measured decay rates deviate significantly from the point-dipole theory, i.e. significantly slower (faster) QD decay dynamics is observed for the direct (inverted) structure compared to the expectations for a dipole emitter. These deviations originate from the mesoscopic nature of QDs implying that the electron and hole wavefunctions are spatially extended, as pictured in Fig.~1b. The experimental data are well explained by our theory, which accounts for the mesoscopic size of the QDs, while we can independently rule out alternative mechanisms, e.g., carrier tunneling, non-radiative processes, light-hole contribution to the dipole moment, or quantum-confined Stark shifts \cite{PhysRevB.70.201308} of the QD energy levels induced by the metallic surface (see Supplementary Information). The deviations from dipole theory are observed only for QDs positioned closer than $\sim 100\ \nano\meter$ to the mirror, which is equal to the length scale of the plasmonic penetration depth into the GaAs substrate. The observed variations in decay rate directly illustrate that the mesoscopic character of the QDs strongly influences the coupling to plasmons and can be employed as a resource to either suppress (direct structure) or promote (inverted structure) the excitation of plasmons.

The modified excitation of plasmons  stems from the mesoscopic dimensions of the QD \cite{PhysRevB.68.161307}.  Here we develop a novel model for spontaneous emission from mesoscopic QDs that includes the  spatial extend and asymmetry of QD wavefunctions. By expanding the interaction to first order around the center of the QD, we obtain the total decay rate
\begin{equation}
\gamma_\mathrm{QD} (z) =  \gamma_\mathrm{pd} (z) + \gamma_\mathrm{me} (z),
\end{equation}
where $\gamma_\mathrm{pd}(z)$ is the familiar point-dipole contribution which depends on the transition dipole moment proportional to $\mu_x=\langle \Psi_\mathrm{h} | \hat{p}_x | \Psi_\mathrm{e} \rangle$, and $\gamma_\mathrm{me}(z)$ is the first-order mesoscopic contribution, which is characterized by the moment $\Lambda_{z,x}=\langle \Psi_\mathrm{h} | \hat{p}_z \hat{x}  | \Psi_\mathrm{e} \rangle$, in the following denoted $\mu$ and $\Lambda$.  Here $\Psi_\mathrm{e}$ ($\Psi_\mathrm{h}$) is the wavefunction of an electron (hole) trapped in the QD, and $\hat{x}$ and $\hat{p}_z$ are position and momentum operators for the $x$ and $z$ directions, respectively. $\Lambda$ is an intrinsic property of the QD and is determined by the size and geometry of the electron and hole wavefunctions.

Due to their mesoscopic dimensions and asymmetric confinement potential QD wavefunctions are extended and asymmetric \cite{PhysRevB.70.201308}, as sketched in Fig.~1b. The mesoscopic decay rate $\gamma_\mathrm{me} (z)$ depends not only on $\Lambda$ but also on the optical field gradient, which is large for plasmonic modes, cf. Fig.~1a. The two contributions to the decay rate in Eq.~(1) combine coherently and therefore can either add or subtract depending on the specific nanophotonic structure surrounding the QD, as was observed for the direct and inverted mirror structures in the data of Fig.~2.  This novel effect has no counterpart in atomic systems where the higher-order interactions between light and atoms are restricted by selection rules, i.e. co-existence of  the first higher-order and dipole transitions is prohibited by the symmetry of the atomic potential. The moment $\Lambda$ contains both electric quadrupolar and magnetic dipolar terms and would in the case of atomic-like emitters vanish on electric-dipole transitions \cite{NovQuadru,OPTEXP_Rukhlenko_Jagadish_2009}. We observe that for QD emitters the higher-order processes can strongly modify the dipole transition due to the mesoscopic nature of the wavefunctions thereby enhancing the light-matter interaction strength significantly.

The novel theory models the experimental data of Fig.~2 very well for both the direct and inverted structure. From the comparison we extract the experimental values  $\Lambda = (9.8 \pm 1.4)\times 10^{-33}\ \kilogram \ \meter^2 \ \rp\second $ ($\Lambda = (\text{-}6.5 \pm 0.8)\times 10^{-33}\ \kilogram \ \meter^2 \ \rp\second $) for the direct (inverted) structure from the experimental value of $\mu$ obtained  using the method of Ref.~\cite{johansen:073303}. The observed change of the sign of $\Lambda$, stems from the opposite orientation of the QDs relative to the plasmonic field and constitutes the tell-tale of the mesoscopic effects.

From our comparison with theory we can extract the rate of excitation of plasmons $\gamma_\mathrm{pl}(z)$, which  should be maximized in  fast quantum plasmonic devices. In Fig.~3a, the extracted $\gamma_\mathrm{pl}(z)$ (see Supplementary Information) is plotted versus distance to the mirror for both the direct and inverted structures. A pronounced difference of the plasmon excitation rate by a factor of eight is observed between the two structures. In contrast, the spontaneous-emission rate of photons $\gamma_\mathrm{ph}(z)$ (Fig.~3b) is similar for the two different structures. These observations can be explained from the fact that plasmon modes give rise to strong electric-field gradients near the metal mirror, thereby enhancing the influence of the mesoscopic QD effects. We note that the extracted mesoscopic and point-dipole contributions to the plasmon excitation rate are of equal magnitude. Therefore the mesoscopic contribution is so pronounced that the applied first-order perturbation theory is pushed to the limit of validity, which could account for the slight difference in the magnitudes of $\Lambda$ between the two datasets. The figure-of-merit of a quantum plasmonic device is the $\beta$-factor expressing the probability that a QD excites a single plasmon: $\beta_\mathrm{pl}(z) = \frac{\gamma_\mathrm{pl}(z)}{\gamma_\mathrm{QD}(z)},$ which is plotted in Fig.~3c. The $\beta$-factor is strongly enhanced due to the mesoscopic effects reaching 40 \% for the inverted sample as opposed to  13 \% for the direct sample where it is suppressed. These observations illustrate the potential of using the intrinsic mesoscopic properties of QDs in combination with careful engineering of the electromagnetic environment to strongly enhance the coupling to plasmons.

We further investigate the potential of mesoscopic QDs for improving plasmon-nanowire single-photon sources \cite{chang:053002,Akimov2007,singlephotontransistor,PhysRevB.78.153111}, see Fig.~4b. For a small wire radius ($r=12.5\ \nano\meter$) only a single strongly confined plasmon exists inducing very strong field gradients, i.e. mesoscopic QD effects are expected to be very pronounced. We note that structures of this size can readily be fabricated by electron beam lithography or chemical synthesis.  We calculate $\gamma_\mathrm{pl}(z)$ versus distance to the nanowire for varying ratios of $\Lambda/\mu$, corresponding to QDs with various amount of mesoscopic character, and for two different orientations of the QD relative to the nanowire. The resulting plasmonic coupling efficiency $\beta_\mathrm{pl}$ is shown in Fig.~4a. Very strong dependencies on both distance and $\Lambda/\mu$ are observed. For a fixed distance of $d=10 \ \nano \meter$ we find that the efficiency for a point-dipole source (i.e. $\Lambda =0$) is $\beta_\mathrm{pl} = 75 \%$. This number can be enhanced substantially to  $\beta_\mathrm{pl} = 92 \%$ assuming the experimental value of $\Lambda /\mu \approx \text{-}10 \ \nano \meter$ for a QD placed near the nanowire, see Fig.~4b. On the other hand the same QD oriented upside-down relative to the nanowire  ($\Lambda/\mu \approx 10 \ \nano \meter$) would only couple weakly to the nanowire with $\beta_\mathrm{pl} < 1 \%$. These results demonstrate the very pronounced effects of including the naturally occurring mesoscopic contribution to the QD decay and that it can be employed for improving the efficiency of plasmonic nanophotonic devices.

We have demonstrated that the interaction between QDs and plasmonic nanostructures can  be understood only by taking the mesoscopic size of the QDs into account. Our findings  are expected to be of relevance also for dielectric nanostrucures, where mesoscopic QD effects are anticipated to be of importance for spontaneous-emission control in photonic crystals \cite{nat04_lodahl_vanmaekelbergh_vos}, dielectric-waveguide single-photon sources \cite{PhysRevLett.101.113903}, and in cavity QED \cite{Hennessy} in particular when employing large QD emitters that currently are intensively investigated for their prospective large oscillator strength \cite{Andreani-PhysRevB.60.13276}. Our conclusions are surprising since the point-dipole approximation has been uncritically adopted in the litterature to describe light-matter interaction between QDs and nanophotonic structures.  Importantly, the mesoscopic effects are very pronounced and may be employed as a resource to enhance light-matter interaction, which is required in a diverse range of scientific fields ranging from quantum information science and quantum computing to energy harvesting devices.

\section*{Supplementary Information}
{\bf Sample preparation.} The semiconductor wafer used for both the direct and inverted sample was grown by molecular beam epitaxy with the following layers from bottom to top: A GaAs substrate, a $50 \ \nano\meter$ AlAs sacrificial layer, a $623 \ \nano \meter$ GaAs buffer, $2.13$ monolayers of InAs, and a $302 \ \nano\meter$ capping layer of GaAs. The InAs layer formed QDs with a density of $250 \ \micro \rpsquare \meter$. Finally, an optically thick ($200 \ \nano\meter$) silver mirror was deposited on the surface of the sample.

For the direct sample a series of terraces with different distance to the underlying QDs were fabricated by UV-lithography and wet etching, for further details see Ref.~\cite{stobbe:155307}. Optical access was provided by selective etching of the AlAs layer and subsequential epitaxial lift-off of the layers above the AlAs and bonding to a sapphire substate. For the inverted sample, the layers above the AlAs were bonded to a PMMA-coated silicon substrate. The sample was transferred to an SU-8-coated sapphire substrate and bonded upside down. After removal of PMMA by oxygen plasma ashering, the terrace fabrication and silver evaporation were carried out as for the direct sample.

Due to the different thicknesses of the buffer and capping layers, the resulting distances to the sapphire substrate are different for the direct and inverted samples. This gives rise to a small difference between the decay rates as a function of distance to the mirror in the two samples, which can be seen by comparing the curves in Fig. 2a and 2b, or in Fig. 3b.

{\bf Measurement setup.} The samples were placed in a closed-cycle cryostat kept at $16\ \kelvin$. The QDs were excited with a Ti:sapphire laser that emits picosecond pulses at a repetition rate of $76 \ \mega \hertz$. The laser was tuned to $1.45\ \electronvolt$, which corresponds to absorption in the wetting layer. The power was adjusted so as to only populate the ground states of the QDs. The spontaneously emitted light was collected with a lens after which is was imaged onto a monochromator where the inhomogeneously broadened spontaneous-emission spectrum from the QDs was spatially dispersed. A thin slit was used to select a narrow band ($2.6\ \milli \electronvolt$) of the spectrum centered at $1.204\ \electronvolt$. This corresponds to a low energy in the inhomogeneously broadened QD spectrum, which together with the weak pumping conditions ($\sim$ 0.1 excitons/QD) ensures that excited states from QDs with lower ground state energy do not contribute to the selected emission. The collected emission was then measured with a fast avalanche photo diode. The resulting decay curves are fitted with a bi-exponential model and the fast decay rate extracted, for further details see Ref.~\cite{stobbe:155307}.

{\bf Decay rate modeling.} We now present the main steps in the calculation of the decay rate of the QD beyond the point-dipole approximation. The minimal-coupling Hamiltonian describing a charged particle interacting with an optical field is given by
\begin{equation}
H({\bf r}, t) = \frac{1}{2 m} [{\bf \hat{p}} - q {\bf A}({\bf r})]^2 + q \phi({\bf r}) + V({\bf r}),
\end{equation}
where $V({\bf r})$ is the Coulomb potential, $q$ is the charge, $m$ is the mass, and ${\bf \hat{p}}$ is the momentum operator of the particle. ${\bf A}({\bf r})$ and $\phi({\bf r})$ are the vector and scalar potentials of the optical field at the position ${\bf r}$ of the particle. We use the generalized Coulomb gauge, which means that $\nabla \cdot (\epsilon({\bf r})  {\bf A} ({\bf r}))=0$. We then introduce raising and lowering operators for both the optical field and the two-level electronic system, switch to the interaction picture, and employ the rotating-wave approximation to arrive at
\begin{widetext}
\begin{equation}
H_I'({\bf r}, t) = -\frac{q}{m} \sum_{l,j} \left(\frac{\hbar}{2 \epsilon_0 \omega_l} \right)^{1/2} \left(e^{i \Delta_l t} \hat{\sigma}_- \hat{a}_l^\dag \langle v| \hat{p}_j f_{l,j}^*({\bf r})|c\rangle + e^{-i \Delta_l t} \hat{\sigma}_+ \hat{a}_l \langle c| \hat{p}_j f_{l,j}({\bf r})|v\rangle \right).
\end{equation}
\end{widetext}
Here $\epsilon_0$ is the permittivity of vacuum, and $| c \rangle$ ($| v \rangle$) is the state of an electron (a hole) in the conduction (valence) band. $f_{l,j}({\bf r})$ denote vector components of mode functions for the optical field, with $l$ indexing the modes and $j$ indexing the three coordinates $x,y,z$. The frequency of the mode function is $\omega_l$, and we define the detuning $\Delta_l = \omega_l - \omega_0$, where $\omega_0$ is the frequency difference between the two electronic states. $\hat{\sigma}_-$ and $\hat{\sigma}_+$ are the raising and lowering operators for the electronic states and $\hat{a}_l^\dag$ creates one photon in mode $l$ while $\hat{a}_l$ annihilates one. We Taylor-expand the field modes to first order around the center of the QD as
\begin{equation}
f_{l,j}({\bf r}) \approx f_{l,j}({\bf r}_0) + \sum_n ({{\bf r - r}_0})_n \left[ \nabla_n f_{l,j}({\bf r}) \right]_{{\bf r = r}_0},
\end{equation}
which gives the following expression for the interaction Hamiltonian
\begin{widetext}
\begin{eqnarray}
H_I'({\bf r_0}, t) &=& -\frac{q}{m} \sum_{j,l,n} \left(\frac{\hbar}{2 \epsilon_0 \omega_l} \right)^{1/2} e^{i \Delta_l t} \hat{\sigma}_- \hat{a}_l^\dag \left[\left(\mu_j + \Lambda_{j,n} \nabla_n \right)f_{l,j}^*({\bf r}) \right]_{{\bf r = r}_0} \nonumber \\
&& -\frac{q}{m} \sum_{j,l,n} \left(\frac{\hbar}{2 \epsilon_0 \omega_l} \right)^{1/2} e^{-i \Delta_l t} \hat{\sigma}_+ \hat{a}_l \left[\left(\mu_j^* + \Lambda_{j,n}^* \nabla_n \right)f_{l,j}({\bf r}) \right]_{{\bf r = r}_0},
\end{eqnarray}
\end{widetext}
where $\mu_j = \langle v | \hat{p}_j |c\rangle$ and $\Lambda_{j,n} = \langle v | \hat{p}_j \hat{r}_n |c\rangle$. In order to find the decay rate of the excited state we use the Schr\"{o}dinger equation in the interaction picture with the state of the system given by $\Psi(t) = c_e(t) |c \rangle |0 \rangle + \sum_l c_l(t) |v \rangle |l \rangle$. The decay of the excited-state population $| c_e(t) |^2 = \exp[-\Gamma t]$ with a rate $\Gamma$ is calculated in the Wigner-Weisskopf approximation. The sum over the mode functions is directly related to the Green's function $ G_{j,j'}({\bf r}, {\bf r} ' ; \omega)$, which describes the environment surrounding the QD. The resulting expression for the decay rate is
\begin{widetext}
\begin{equation}
\Gamma({\bf r}_0, \omega) = \frac{2 q^2}{ c^2 \epsilon_0 \hbar m^2} \sum_{j,j',n,n'} \left[ \mu_j
+ \Lambda_{j,n} \nabla_{n} \right] \left[ \mu_{j'}^*
+ \Lambda_{j',n'}^* \nabla_{n'}'\right]
 \mathrm{Im}( G_{j,j'}({\bf r}, {\bf r} ' ; \omega))     \Big|_{{\bf r} = {\bf r} ' = {\bf r}_0}.
\end{equation}
\end{widetext}
The decay rate obtained beyond the dipole approximation (Eq. (S5)) can be cast in the form of an integral over the length of the in-plane wavevector $k_{\parallel}$ and divided into parts associated with freely propagating photons, and bound modes \cite{kalkman:075317}. In the case of a metallic mirror we can use the rotational symmetry to simplify the expressions, using $2 \pi \sqrt{\epsilon} \lambda_0^{-1} =k_\mathrm{dielectric} = \sqrt{k_z^2+k_{\parallel}^2}$ with $\epsilon$ being the permittivity of the dielectric and $\lambda_0$ the wavelength of light in vacuum. For $k_{\parallel} \le k_\mathrm{dielectric}$ the modes have real-valued $k_z$ and propagate as free photons, while for $k_{\parallel} > k_\mathrm{dielectric}$ the modes have imaginary-valued $k_z$ and are bound to the interface. The bound modes can be subdivided into propagating plasmons that fulfill $k_{\parallel} \approx k_\mathrm{pl}$ and lossy modes that do not.

The wavefunctions for electron and hole states are decomposed into envelope parts and Bloch parts, which describe the wavefunctions on the scale of the QD and on the length scale of a crystal unit cell, respectively. The envelope functions are symmetric in the $x$ and $y$ directions and asymmetric in the $z$ direction due to the symmetry of self-assembled QDs. The Bloch function for the electron in the conduction band has even symmetry in all directions, while the Bloch function for the heavy hole is a superposition of two states, the first one even in $x$,  $z$ and odd in $y$, and the second state is even in $y$, $z$ and odd in $x$. These symmetries of the wave functions give $|\mu_x| = |\mu_y| = \mu$,  $\mu_z = 0$, and furthermore imply that the mesoscopic size-induced moments $\Lambda_{j,n}$ have $\Lambda_{x,x} = \Lambda_{x,y} = \Lambda_{y,x} = \Lambda_{y,y} = \Lambda_{z,z} = 0$, $\Lambda_1 = \Lambda_{z,x} = \Lambda_{z,y}$ and, $\Lambda_2=\Lambda_{x,z} = \Lambda_{y,z}$.
The mesoscopic moments $\Lambda_1$ and $\Lambda_2$ couple to two different polarizations of the plasmonic field. The ratio of the field strengths of these two polarizations is given by $|\mathrm{Re}[\epsilon_\mathrm{Ag}]/\epsilon_\mathrm{GaAs}| = 4.3$. Due to this large ratio we choose to neglect $\Lambda_2$, that couples to the weaker field, and define the moment $\Lambda$ used in the main text as $\Lambda = \Lambda_1$. As a result we have only one parameter describing the effect of the higher-order moments in our decay rate model:
\begin{equation}
\bar{\mu} = \mu \left( \begin{array}{c}
1 \\
i \\
0
\end{array} \right), \hspace{1 cm} \bar{\Lambda} = \Lambda \left( \begin{array}{ccc}
0 & 0 & 0 \\
0 & 0 & 0 \\
1 & i & 0
\end{array} \right).
\end{equation}
 We have here chosen the Bloch wavefunction for the heavy hole given by the plus sign in $|u_\mathrm{hh}\rangle = |u_x \rangle \pm i |u_y \rangle$ corresponding to a particular exciton spin state. We note that the decay dynamics for the silver mirror geometry considered here does not depend on which of the in-plane heavy hole states that is excited.

We note that the first-order moment $\Lambda$ couples to the gradient of the Green's function while the point-dipole moment $\mu$ couples to its size, see (Eq. (S5)). For coupling to the plasmonic modes this corresponds to a scaling proportional to $i k_\mathrm{pl} \Lambda  e^{i k_z z}$ for the first-order contribution, and $\mu e^{i k_z z}$ for the zeroth-order. The breakdown of the dipole approximation is thus determined by the ratio of the zeroth and first-order contributions, which is independent of the distance $z$ to the silver mirror.

We normalize the QD decay rate relative to the decay rate in a homogenous medium leading to an expression, which depends only on the ratio $\Lambda/\mu$. The intrinsic radiative decay rate $\gamma_\mathrm{rad}$ and non-radiative decay rate $\gamma_\mathrm{nr}$ of the QDs in a homogenous medium are extracted from measurements of the decay rate as function of distance to a GaAs/air interface \cite{johansen:073303} using the inverted sample before silver was evaporated. These measurements are described well by the point-dipole term only and we extract $\gamma_\mathrm{rad} = 0.88 \ \nano \rp\second$ and $\gamma_\mathrm{nr} = 0.19 \ \nano \rp\second$.

Figure 3 in the main text is derived from the measured decay rates $\gamma_\mathrm{QD}(z)$, $\gamma_\mathrm{rad}$, and $\gamma_\mathrm{nr}$,
and the calculated decay rates discussed above $\gamma_\mathrm{pl}(z)$, $\gamma_\mathrm{ph}(z)$, and $\gamma_\mathrm{ls}(z)$.
We can extract the data points in Fig. 3a, as
\begin{equation}
\gamma_\mathrm{pl}(z) = \gamma_\mathrm{QD}(z) - \gamma_\mathrm{ph}(z) - \gamma_\mathrm{ls}(z) - \gamma_\mathrm{nr}.
\end{equation}
Figure 3b is derived in the same manner and Fig. 3c is derived as $\beta_\mathrm{pl}(z) = \gamma_\mathrm{pl}(z) / \gamma_\mathrm{QD}(z)$.

For the nanowire calculation the coupling efficiency to the plasmonic mode is  calculated for an emitter with a quantum efficiency of unity, i.e.  $\gamma_\mathrm{nr}= 0$. The decay rate into photons $\gamma_\mathrm{ph}$ is assumed constant and equal to the value in homogenous GaAs. The plasmon mode was calculated by a finite-element method \cite{PhysRevB.81.125431} implemented in COMSOL and the corresponding Green's function constructed. This is a feasible approach since only a single plasmon mode exists for the studied wire of radius $r = 12.5 \ \nano\meter$. Based on these simulations, $\gamma_\mathrm{pl}$ is extracted including the effect of a spatially extended QD with $\Lambda \ne 0$. The decay rate into lossy modes $\gamma_\mathrm{ls}$ is calculated from an analytical expression  for a point dipole near a nanowire \cite{PhysRevB.76.035420,PhysRevA.69.013812}.   Here we have modeled a dipole-moment that is oriented at 45 degrees to both the azimuthal and parallel direction of the wire (see Fig.~4 of the main text), to give an average effect on the modified decay efficiencies. Stronger (weaker) modifications of the decay rate result for QD dipole moments oriented parallelly (azimuthally) to the nanowire. All the calculations are performed at $\lambda_0 = 1030 \ \nano \meter$.

{\bf Ruling out alternative mechanisms for the observed effects.}
When observing a new physical effect it is essential to be able to rule out that the observations could be dominated by alternative mechanisms. The key tell-tale for the mesoscopic effects signifying the breakdown of the dipole approximation on a dipole allowed transition is the change from enhancement to suppression of the rate when reversing the QD orientation. In the following we explicitly explain why alternative mechanisms can be ruled out.

The excitation in a QD can be lost by tunneling of either the trapped electron or hole out of the QD. Such non-radiative processes may be enhanced near surfaces \cite{johansen:073303}, but can only increase the measured total decay rate. Therefore, non-radiative processes cannot explain the measured suppression of the decay rate observed for the direct sample and can be ruled out.

It has been predicted that the hole wavefunction partly resides in the light-hole band \cite{PRB_Bester_Zunger_2003}, which results in an out-of-plane component of the dipole-moment. However, also a dipole oriented out-of-plane would radiate identically in the direct and inverted structure and thus would not explain our observations.

The decay rate of a QD treated in the dipole approximation is proportional to the square of the overlap of the envelope functions for the confined electron and hole. By applying an electrical field over the QD along the growth direction it
is possible to change the aforementioned overlap and thereby the decay rate of the QD, which is a direct consequence of the quantum-confined Stark shift \cite{PhysRevB.70.201308}. An unavoidable impurity background doping of the semiconductor along with the silver mirror forms a Schottky barrier with such a build-in electrical field. The background ion impurity density of the used wafer has been measured to be $N = 4.3 \times 10^{21}\ \meter ^{-3}$ from mobility measurements on a two-dimensional electron gas grown in the molecular beam epitaxy chamber. From this number we can calculate the typical length scale of the surface induced electric field into the GaAs as \cite{Book_ThePhysicsofLowDimensionalSemiconductors} $d = \sqrt{2 V_b \epsilon_0 \epsilon_d/(e^2 N)} = 462-580\ \nano\meter$ , where $V_b = 0.71 - 1.12\ \electronvolt$ is the barrier height \cite{JEM_Arulkumaran_Ramasamy_1995}, $e$ is the elementary charge, and $\epsilon_d$ is the permittivity of GaAs. Thus any resulting surface electric field would extend 5-6 times further than the length over which we observe deviations from dipole theory, and we can therefore rule out the Stark effect as explanation of our data.   The conclusion is confirmed by calculating the resultant electric field ($0-24\ \kilo\volt/\centi\meter$) in our structure. Previous experiments with applied electric fields have shown that no significant change in the decay rate occurs in this range \cite{alen:045319}. Notably no enhancement of the decay rate is observed, which is the case for the inverted structure in our experiment. Finally, the typical dominant impurities in GaAs are carbon defects giving rise to a p-type GaAs. From measurements of the static dipole moment of QDs \cite{PhysRevB.70.201308} it is found that the hole is situated above the electron for zero applied electric field, which together with a Schottky barrier with p-type GaAs results in a reduction of the decay rate in the inverted structure in conflict with our measurements.

\subsection*{Acknowledgements}
We would like to thank Yuntian Chen for supplying numerical code for the finite element calculations and Martijn Wubs and J\o rn M. Hvam for careful comments on the manuscript. We gratefully acknowledge financial support from the Villum Kann Rasmussen Foundation, The Danish Council for Independent Research (Natural Sciences and Technology and Production Sciences), and the Danish National Research Foundation.

\subsection*{Author Contributions}
M.L.A. and S.S. fabricated the samples and developed the concepts. M.L.A. carried out the experiment, analyzed the data, and implemented the theoretical model. A.S.S. suggested the use of the inverted structure and supported the theoretical work. M.L.A. and P.L. wrote the paper. P.L. initiated and supervised the project. All authors provided detailed comments on the manuscript.

\section*{Figure legends}

{\bf Figure 1 \textbar\  Mesoscopic QDs in plasmonic nanostructures.} {\bf a}, Sketch of the studied system. A QD (green trapezoid) is placed a distance $z$ below a metal mirror. The lateral extension of a QD emitting at $1.2 \ \electronvolt$ is typically $a=20 \ \nano \meter$. The plasmon wavelength is $\lambda_\mathrm{pl} = 262 \ \nano \meter$ (figure is not to scale). The field amplitude of the plasmon decays exponentially away from the interface with a change of the electric field over the extension of the QD. The arrow over $\mu$ indicates the orientation of the point-dipole moment and the arrows over $\Lambda$ the orientation of the first order mesoscopic moment. {\bf b}, Boundaries of a QD (green frame) with the spatial extension of electron (blue) and hole (red) wavefunctions indicated inside. {\bf c}, Sketch of a QD placed near a metallic structure. The QD can decay by emitting a photon ($\gamma_\mathrm{ph}$), by exciting a propagating plasmon ($\gamma_\mathrm{pl}$), by coupling to lossy modes in the metal ($\gamma_\mathrm{ls}$), or by intrinsic non-radiative recombination ($\gamma_\mathrm{nr}$) (not shown) .

{\bf Figure 2 \textbar\ Observation of the breakdown of the dipole approximation.} Measured decay rates of QDs  as a function of distance to the silver mirror for the direct ({\bf a}) and inverted ({\bf b}) structure at a wavelength of  $\lambda = 1030 \ \nano \meter$. The dashed curves are the predicted variation for a point-dipole emitter.  The solid curves show the theory for a mesoscopic emitter, which are found to match the experimental data very well. The insets show the orientation of the QDs relative to the silver mirrors for the two structures. The error bars on both rates and positions represent one standard deviation and are deduced from repeated measurements.

{\bf Figure 3 \textbar\ Influence of mesoscopic effects on decay rates.} The decay rate of QDs to plasmons $\gamma_\mathrm{pl}$ ({\bf a}) and photons $\gamma_\mathrm{ph}$ ({\bf b}), and the plasmon generation efficiency $\beta_\mathrm{pl}$ ({\bf c}) as a function of distance to the silver mirror. Red \ding{116} and red curves are experimental data and theory for the inverted structure, while the blue \ding{115} and blue curves are the equivalent for the direct structure. The error bars on both rates and positions represent one standard deviation, as deduced from repeated measurements.

{\bf Figure 4 \textbar\ Efficiency of nanoplasmonic single-photon source with mesoscopic emitter.} {\bf a}, Plasmon generation efficiency  $\beta_\mathrm{pl}$ for a QD in GaAs near a silver nanowire with radius $r=12.5\ \nano \meter$ for varying distance $d$ and $\Lambda/\mu$. We have used a constant $\gamma_\mathrm{ph}$ and included $\gamma_\mathrm{ls}$ for a point dipole, see Supplementary Information. The dotted lines indicate $\Lambda/\mu = \pm 10 \ \nano \meter$, which are representative for the experiment. {\bf b}, Sketch of a nanowire with $r=12.5\ \nano \meter$ and mesoscopic QDs positioned a distance $d$ from the surface.  The dipole moment of the modeled QD is oriented at $45^\circ$ to both the parallel ($\hat{r}_\parallel$) and azimuthal ($\hat{\phi}$) directions. Negative (positive) $\Lambda/\mu$ corresponds to a QD positioned above (below) the nanowire.

\clearpage

\begin{figure}[htp]
\centering
\includegraphics[width=\columnwidth]{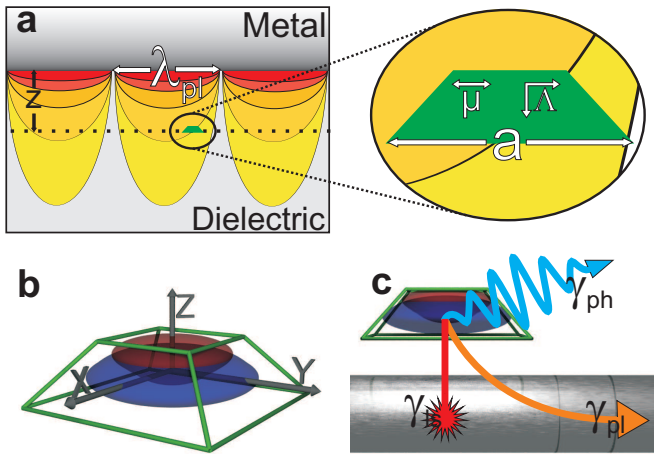}
\caption{{\bf Mesoscopic QDs in plasmonic nanostructures.}}
\end{figure}

\begin{figure}[htp]
\centering
\includegraphics[width=\columnwidth]{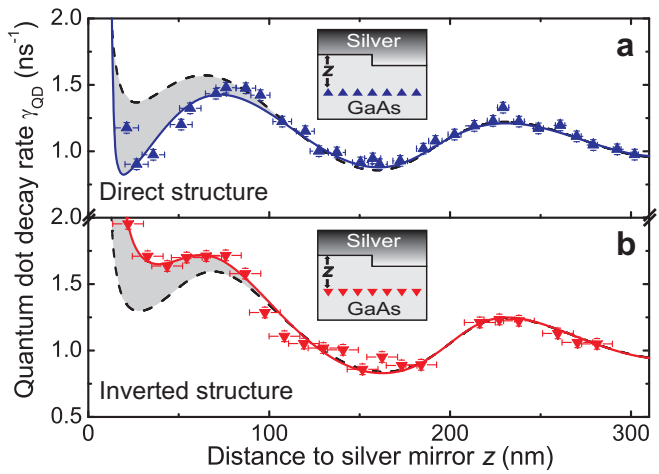}
\caption{{\bf Observation of the breakdown of the dipole approximation.}}
\end{figure}

\begin{figure}[htp]
\centering
\includegraphics[width=\columnwidth]{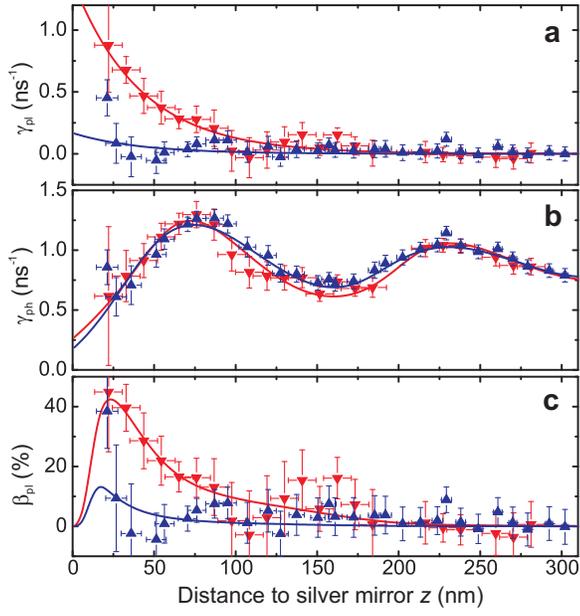}
\caption{{\bf  Influence of mesoscopic effects on decay rates.}}
\end{figure}

\begin{figure}[htp]
\centering
\includegraphics[width=\columnwidth]{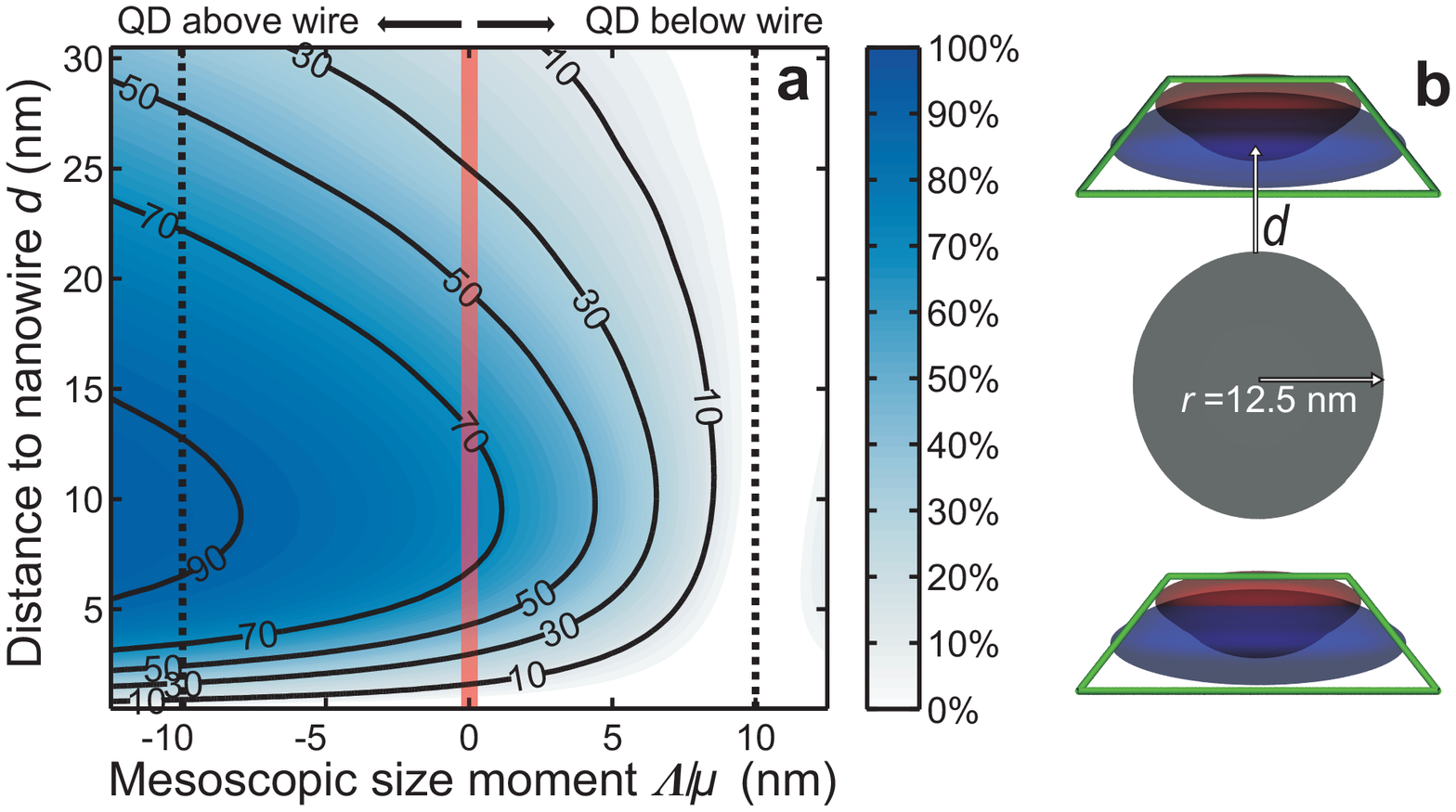}
\caption{{\bf  Efficiency of nanoplasmonic single-photon source with mesoscopic emitter.}}
\end{figure}


\begin{thebibliography}{10}

\bibitem{nmat_Polman}
Atwater, H.~A. \& Polman, A. Plasmonics for improved photovoltaic devices
  \textit{Nature Materials} 9 205--213 (2010).

\bibitem{Nozik2002115}
Nozik, A.~J. Quantum dot solar cells \textit{Physica E: Low-dimensional Systems
  and Nanostructures} 14 115 -- 120 (2002).

\bibitem{PRL_Fattal_Yamamoto_2004}
Fattal, D., Diamanti, E., Inoue, K. \& Yamamoto, Y. Quantum teleportation with
  a quantum dot single photon source \textit{Phys. Rev. Lett.} 92 037904
  (2004).

\bibitem{Nature_Ladd}
Ladd, T.~D. \textit{et~al.} Quantum computers \textit{Nature} 464 45--53
  (2010).

\bibitem{P.Michler12222000}
Michler, P. \textit{et~al.} A quantum dot single-photon turnstile device
  \textit{Science} 290 2282--2285 (2000).

\bibitem{PhysRevLett.90.206803}
Taylor, J.~M., Marcus, C.~M. \& Lukin, M.~D. Long-lived memory for mesoscopic
  quantum bits \textit{Phys. Rev. Lett.} 90 206803 (2003).

\bibitem{PRA_Loss_DiVincenzo_1998}
Loss, D. \& DiVincenzo, D.~P. Quantum computation with quantum dots
  \textit{Phys. Rev. A} 57 120--126 (1998).

\bibitem{Hennessy}
Hennessy, K. \textit{et~al.} Quantum nature of a strongly coupled single
  quantum dot{\textendash}cavity system \textit{Nature} 445 896--899 (2007).

\bibitem{PhysRevLett.101.113903}
Lund-Hansen, T. \textit{et~al.} Experimental realization of highly efficient
  broadband coupling of single quantum dots to a photonic crystal waveguide
  \textit{Phys. Rev. Lett.} 101 113903 (2008).

\bibitem{chang:053002}
Chang, D.~E., S{\o}rensen, A.~S., Hemmer, P.~R. \& Lukin, M.~D. Quantum optics
  with surface plasmons \textit{Phys. Rev. Lett.} 97 053002 (2006).

\bibitem{Akimov2007}
Akimov, A.~V. \textit{et~al.} {Generation of single optical plasmons in
  metallic nanowires coupled to quantum dots} \textit{Nature} 450 402--406
  (2007).

\bibitem{singlephotontransistor}
Chang, D.~E., S{\o}rensen, A.~S., Demler, E.~A. \& Lukin, M.~D. A single-photon
  transistor using nanoscale surface plasmons \textit{Nature Physics} 3
  807--812 (2007).

\bibitem{PhysRevB.78.153111}
Jun, Y.~C., Kekatpure, R.~D., White, J.~S. \& Brongersma, M.~L. Nonresonant
  enhancement of spontaneous emission in metal-dielectric-metal plasmon
  waveguide structures \textit{Phys. Rev. B} 78 153111 (2008).

\bibitem{Springer_Schmidt}
Schmidt, O.~G. (Ed.) \textit{Lateral Alignment of Epitaxial Quantum Dots}
  (Springer, Berlin, 2007).

\bibitem{O'brien09}
O'Brien, J.~L., Furusawa, A. \& Vuckovic, J. Photonic quantum technologies
  \textit{Nature Photonics} 3 687--695 (2009).

\bibitem{strauf:127404}
Strauf, S. \textit{et~al.} Self-tuned quantum dot gain in photonic crystal
  lasers \textit{Phys. Rev. Lett.} 96 127404 (2006).

\bibitem{nphoton-plasmonicsreview}
Gramotnev, D.~K. \& Bozhevolnyi, S.~I. Plasmonics beyond the diffraction limit
  \textit{Nature Photonics} 4 83--91 (2010).

\bibitem{nmat-brongesma}
Schuller, J.~A. \textit{et~al.} Plasmonics for extreme light concentration and
  manipulation \textit{Nature Materials} 9 193--204 (2010).

\bibitem{naturephys_spp_duality}
Kolesov, R. \textit{et~al.} Wave particle duality of single surface plasmon
  polaritons \textit{Nature Physics} 5 470--474 (2009).

\bibitem{spaser_first}
Bergman, D.~J. \& Stockman, M.~I. Surface plasmon amplification by stimulated
  emission of radiation: Quantum generation of coherent surface plasmons in
  nanosystems \textit{Phys. Rev. Lett.} 90 027402 (2003).

\bibitem{Noginov2009}
Noginov, M.~A. \textit{et~al.} Demonstration of a spaser-based nanolaser
  \textit{Nature} 460 1110--1112 (2009).

\bibitem{nat04_lodahl_vanmaekelbergh_vos}
Lodahl, P. \textit{et~al.} Controlling the dynamics of spontaneous emission
  from quantum dots by photonic crystals \textit{Nature} 430 654--657 (2004).

\bibitem{JLumin_Drexhage_1970}
Drexhage, K.~H. Influence of a dielectric interface on fluorescence decay time
  \textit{Journal of Luminescence} 1-2 693 -- 701 (1970).

\bibitem{PhysRevB.70.201308}
Finley, J.~J. \textit{et~al.} Quantum-confined Stark shifts of charged exciton
  complexes in quantum dots \textit{Phys. Rev. B} 70 201308 (2004).

\bibitem{PhysRevB.68.161307}
Jun~Ahn, K. \& Knorr, A. Radiative lifetime of quantum confined excitons near
  interfaces \textit{Phys. Rev. B} 68 161307 (2003).

\bibitem{NovQuadru}
Zurita-S\'{a}nchez, J.~R. \& Novotny, L. Multipolar interband absorption in a
  semiconductor quantum dot. I. Electric quadrupole enhancement \textit{J. Opt.
  Soc. Am. B} 19 1355--1362 (2002).

\bibitem{OPTEXP_Rukhlenko_Jagadish_2009}
Rukhlenko, I.~D. \textit{et~al.} Spontaneous emission of guided polaritons by
  quantum dot coupled to metallic nanowire: Beyond the dipole approximation
  \textit{Opt. Express} 17 17570--17581 (2009).

\bibitem{johansen:073303}
Johansen, J. \textit{et~al.} Size dependence of the wavefunction of
  self-assembled InAs quantum dots from time-resolved optical measurements
  \textit{Phys. Rev. B.} 77 073303 (2008).

\bibitem{Andreani-PhysRevB.60.13276}
Andreani, L.~C., Panzarini, G. \& G\'erard, J.-M. Strong-coupling regime for
  quantum boxes in pillar microcavities: Theory \textit{Phys. Rev. B} 60
  13276--13279 (1999).

\bibitem{stobbe:155307}
Stobbe, S., Johansen, J., Kristensen, P.~T., Hvam, J.~M. \& Lodahl, P.
  Frequency dependence of the radiative decay rate of excitons in
  self-assembled quantum dots: Experiment and theory \textit{Phys. Rev. B} 80
  155307 (2009).

\bibitem{kalkman:075317}
Kalkman, J., Gersen, H., Kuipers, L. \& Polman, A. Excitation of surface
  plasmons at a SiO$_2$/Ag interface by silicon quantum dots: Experiment and
  theory \textit{Phys. Rev. B} 73 075317 (2006).

\bibitem{PhysRevB.81.125431}
Chen, Y., Nielsen, T.~R., Gregersen, N., Lodahl, P. \& M\o{}rk, J.
  Finite-element modeling of spontaneous emission of a quantum emitter at
  nanoscale proximity to plasmonic waveguides \textit{Phys. Rev. B} 81 125431
  (2010).

\bibitem{PhysRevB.76.035420}
Chang, D.~E., S{\o}rensen, A.~S., Hemmer, P.~R. \& Lukin, M.~D. Strong coupling
  of single emitters to surface plasmons \textit{Phys. Rev. B} 76 035420
  (2007).

\bibitem{PhysRevA.69.013812}
Klimov, V.~V. \& Ducloy, M. Spontaneous emission rate of an excited atom placed
  near a nanofiber \textit{Phys. Rev. A} 69 013812 (2004).

\bibitem{PRB_Bester_Zunger_2003}
Bester, G., Nair, S. \& Zunger, A. Pseudopotential calculation of the excitonic
  fine structure of million-atom self-assembled In$_{1-x}$Ga$_{x}$As/GaAs quantum
  dots \textit{Phys. Rev. B} 67 161306 (2003).

\bibitem{Book_ThePhysicsofLowDimensionalSemiconductors}
Davies, J.~H. \textit{The Physics of Low-Dimensional Semiconductors} (Cambridge
  University Press, 1998).

\bibitem{JEM_Arulkumaran_Ramasamy_1995}
Arulkumaran, S. \textit{et~al.} Investigations on Au, Ag, and Al Schottky
  diodes on liquid encapsulated Czochralski grown n-GaAs<100> \textit{Journal
  of Electronic Materials} 24 813--817 (1995).

\bibitem{alen:045319}
Al\'{e}n, B. \textit{et~al.} Oscillator strength reduction induced by external
  electric fields in self-assembled quantum dots and rings \textit{Physical
  Review B (Condensed Matter and Materials Physics)} 75 045319 (2007).

\end{thebibliography}
\end{document}